# Applications of hyperbolic metamaterial substrates

Yu Guo, Ward Newman, Cristian L. Cortes and Zubin Jacob*
Department of Electrical and Computer Engineering,
University of Alberta, Edmonton, AB T6G 2V4 Canada
*zjacob@ualberta.ca

Abstract: We review the properties of hyperbolic metamaterials and show that they are promising candidates as substrates for nano-imaging, nano-sensing, fluorescence engineering and for controlling thermal emission. Hyperbolic metamaterials can support unique bulk modes, tunable surface plasmon polaritons as well as surface hyperbolic states (Dyakonov plasmons) that can be used for a variety of applications. We compare the effective medium predictions with practical realizations of hyperbolic metamaterials to show their potential for radiative decay engineering, bio-imaging, sub-surface sensing, meta-plasmonics and super-Planckian thermal emission.

1. Introduction

Metamaterial technologies have matured over the past decade for a variety of applications such as super-resolution imaging [1,2], cloaking [3] and perfect absorption [4]. Various classes of metamaterials have emerged that show exotic electromagnetic properties like negative index [5], optical magnetism [6] , giant chirality [7–9], epsilon-near-zero [10], bi-anisotropy [11] and spatial dispersion [12] among many others. The central guiding principle in all the metamaterials consists of fabricating a medium composed of unit cells far below the size of the wavelength. The unique resonances of the unit cell based on its structure and material composition as well as coupling between the cells lead to a designed macroscopic electromagnetic response.

One class of artificial media which has received a lot of attention are hyperbolic metamaterials [13–15]. They derive their name from the unique form of the isofrequency curve which is hyperbolic instead of circular as in conventional dielectrics. The reason for their widespread interest is due to the relative ease of nanofabrication, broadband non-resonant response, wavelength tunability, bulk three dimensional response and high figure of merit [16]. Hyperbolic metamaterials (HMMs) can be used for a variety of applications from negative index waveguides [13], sub-diffraction photonic funnels [17] to nanoscale resonators [18]. In the visible and near-infrared wavelength regions HMMs are the most promising artificial media for practical applications [19].

In this article, we describe the potential of hyperbolic metamaterial substrates for five distinct applications 1) Fluorescence engineering [20–22] 2) Nano-imaging [23–25] 3) Subsurface sensing [26] 4) Dyakonov plasmons [27,28] 5) Super-Planckian thermal emission [29,30]. Our work presents a unified view for these distinct applications and elucidates many key design principles useful to experimentalists and theorists. We focus on the physical origin of hyperbolic behavior in various practical realizations and compare the device performances to theoretical idealizations. We expect this work to provide an overview and starting point for varied practical applications of hyperbolic metamaterials.

2. Hyperbolic Metamaterials

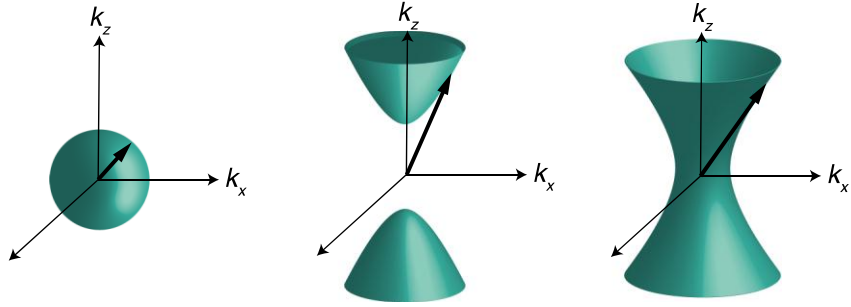

Figure 1: k-space topology. a) The isofrequency contour for an isotropic dielectric is a sphere. For extraordinary waves in an extremely anisotropic uniaxial medium, the isofrequency contour becomes a hyperboloid which supports waves with unbounded wavevectors, in stark contrast to an isotropic medium. (b) A type 1 HMM has one component of the dielectric tensor negative ($\varepsilon_{xx} = \varepsilon_{yy} > 0$ and $\varepsilon_{zz} < 0$) and supports low-$k$ and high-$k$ waves. (c) A type 2 HMM has two components of the dielectric tensor negative ($\varepsilon_{xx} = \varepsilon_{yy} < 0$ and $\varepsilon_{zz} > 0$) and only supports high-$k$ waves.

HMMs can be considered as uniaxial meta-crystals with an extremely anisotropic dielectric tensor, $\ddot{\varepsilon} = diag\left[\varepsilon_{xx}, \varepsilon_{yy}, \varepsilon_{zz}\right]$ such that $\varepsilon_{xx} = \varepsilon_{yy}$ and $\varepsilon_{zz} \bullet \varepsilon_{xx} < 0$. The properties of HMMs are best understood by studying the isofrequency surface of extraordinary waves in this medium

$$\frac{\left(k_x^2 + k_y^2\right)}{\varepsilon_{zz}} + \frac{k_z^2}{\varepsilon_{xx}} = \left(\frac{\omega}{c}\right)^2 \tag{1}$$

The above equation represents a hyperboloid when $\varepsilon_{zz} \bullet \varepsilon_{xx} < 0$ which is an open surface in stark contrast to the closed spherical dispersion in an isotropic medium. The immediate physical consequence of this dispersion relation is the existence of propagating waves with large wavevectors known as high-$k$ waves which are evanescent in conventional media. Multiple device applications and physical phenomena in hyperbolic metamaterials are related to the properties of these high-$k$ waves. It was recently proposed that these states cause a broadband divergence in the photonic density of states, the physical quantity governing various phenomena such as spontaneous and thermal emission [20,31–33]. This prediction led to multiple experimental [21,34] and computational efforts [35,36] to verify the predictions and explore applications of this phenomenon.

We now introduce nomenclature to classify the two types of hyperbolic metamaterials based on the number of components of the dielectric tensor which are negative [19]. Note that if all three components are negative, we have an effective metal and propagating waves are not allowed in such a medium.

- Type I: If there is only one component negative i.e. $\varepsilon_{zz}<0$ in the tensor, then we term such metamaterials as type I HMM. They have low loss because of their predominantly dielectric nature but are difficult to achieve in practice.
- Type II: If there are two components in the dielectric tensor which are negative i.e. $\varepsilon_{xx}=\varepsilon_{yy}<0$, we term them as type II HMMs. They have higher loss and high impedance mismatch with vacuum due to their predominantly metallic nature.

Figure 1 shows the isofrequency surfaces for an isotropic dielectric (i.e. glass) and for a type 1 and type 2 HMM.

3. 1D and 2D realizations

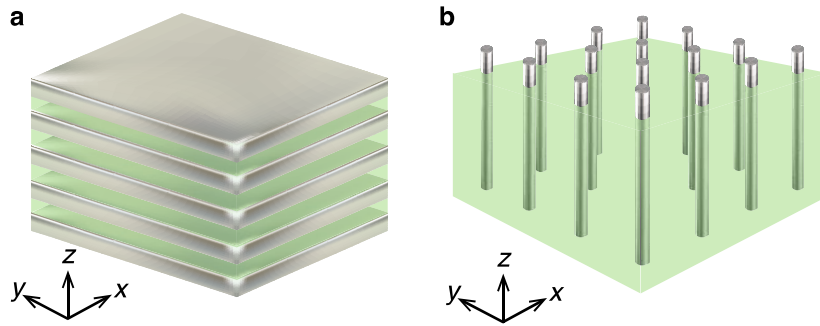

Figure 2: (a) Multilayer realization of hyperbolic metamaterials consisting of alternating subwavelength layers of metal and dielectric (b) metal nanorod realization in a dielectric host matrix

There are two prominent methods to engineer practical hyperbolic media. The first consists of alternating layers of metal and dielectric with the layer thicknesses far below the size of the wavelength. The second approach consists of metal nanorods in a dielectric host such as porous anodic alumina (AAO). Figure 2 is a schematic illustration of these two approaches. Both these approaches achieve the desired extremely anisotropic response according to Maxwell-Garnett effective medium theory [16,37–40] . It is important to note that effective medium theory predicts the desired response in a broad spectral bandwidth because of its non-resonant nature. This is crucial since absorption in resonant metamaterials are a major detriment to practical applications.

- Material systems

The response of the hyperbolic metamaterial can be tuned by the choice of constituent metal and dielectric, and their relative volume ratios.

    o UV and visible: Silver is the ideal choice of metal due to its low losses. Alumina ($Al_2O_3$) is a compatible dielectric in the UV range but at higher wavelengths, the large negative real part of the metallic dielectric constant requires a shift to high index dielectrics. Titanium dioxide is an excellent candidate due to its large index, opening the possibility of impedance matching with vacuum [41].
    o Near-IR: None of the conventional plasmonic metals such as gold or silver are good candidates for hyperbolic metamaterials at near-IR wavelengths. This is because far below the plasma frequency, metals are extremely reflective and lead to high impedance mismatch with surrounding media. The recently developed alternative plasmonic metals based on oxides and nitrides are ideal for applications in hyperbolic metamaterials since their plasma frequency can be tuned to lie in the near-IR [42].
    o Mid-IR: At mid-infrared wavelengths, doped semiconductors can act as the metallic building block for hyperbolic metamaterials [16]. Another option is the use of phonon-polaritonic metals such as silicon carbide which have their Reststrahelen band in the mid-IR range [43].

- What is the origin of the high-k modes?

We now analyze the physical origin of the high-$k$ modes in a practical realization of hyperbolic metamaterials. Due to the metallic building block needed to achieve a negative dielectric constant in one direction, HMMs support bulk plasmon-polaritonic or phonon-polaritonic modes. Thus high-k modes of HMM can be considered as engineered bulk polaritonic modes which owe their large momentum to light-matter coupling. In figure 3, we contrast an incident evanescent wave on a conventional dielectric as opposed to a hyperbolic metamaterial. We take negligible losses to clarify the origin of the high-$k$ states. The evanescent wave decays in a simple dielectric but couples to a high-$k$ propagating wave in the hyperbolic metamaterial. We consider a practical multilayer semiconductor realization of the hyperbolic metamaterial [16]. The high-$k$ mode is seen to arise due to the coupling between the surface-plasmon-polaritons on each of the interfaces. Thus the high-$k$ modes are in fact the bloch modes due to the coupled surface plasmon polaritons on the metal-dielectric multilayer superlattice [12].

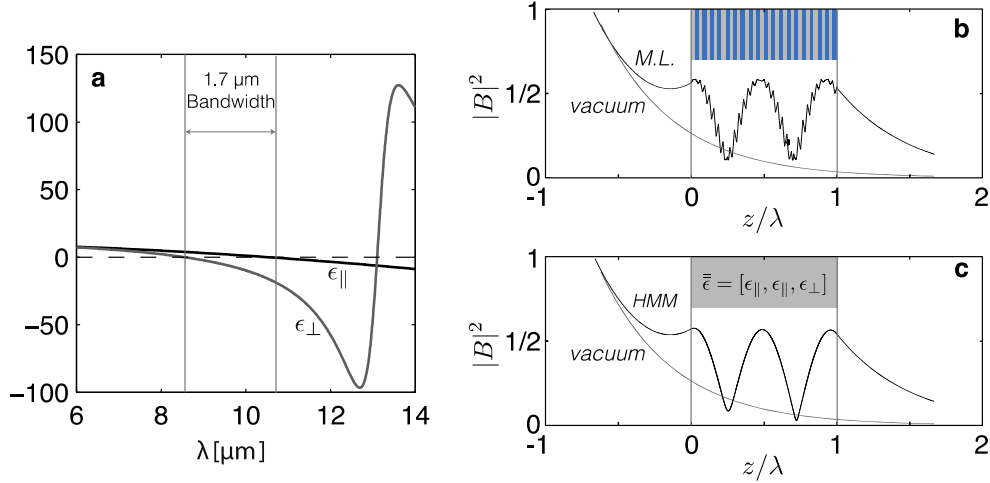

Figure 3: (a) Dispersion of the dielectric constants in a practical multilayer semiconductor realization of the HMM. Note the broadband region in which type I HMM response is achieved (b) Exact numerical calculation (neglecting loss) in the multilayer structure showing the bloch high-*k* mode and coupled surface plasmons at the interfaces (c) Evanescent wave incident on an effective medium slab couples to the high-*k* mode and is transmitted. In contrast, the evanescent wave decays away in vacuum.

- Validity of Effective medium theory

We now consider the transmission of evanescent waves through a practical multilayer structure to determine the largest wavevector that can be transmitted/supported by the hyperbolic metamaterial. In the effective medium limit, waves with infinitely large wavevectors can be transmitted in the entire bandwidth of hyperbolic dispersion. However, this is never true in reality since the size of the unit cell places an upper cut off to the largest wavevector that can be transmitted through the structure. Waves with wavevectors comparable to the inverse unit cell size lie at the edge of the brillouin zone of the periodic lattice. They do not perceive the metamaterial in the effective medium limit but in fact start bragg scattering. This is evident in the comparison shown in figure 4 which considers a Ag/TiO2 multilayer HMM stack both in the effective medium limit and a practical multilayer realization. Our transfer matrix simulations take into account the absorption, dispersion as well as the size of the unit cell. It is seen that high-*k* waves are transmitted through the multilayer structure in excellent agreement with the effective medium theory prediction. The practical realization however has a cut off to the largest wavevector that can be transmitted which is related to the unit cell size of the HMM.

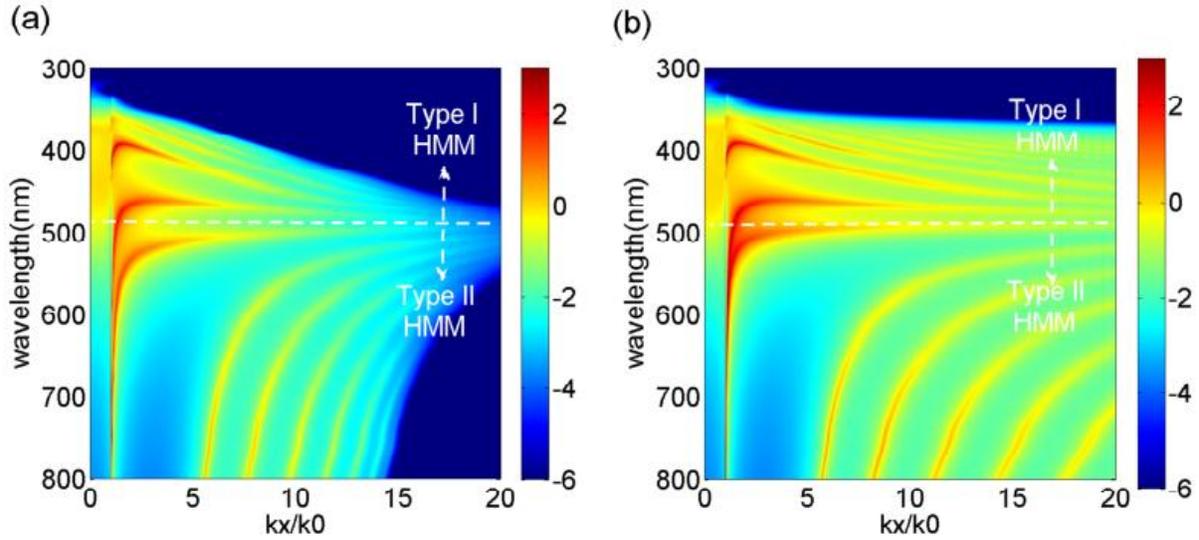

Figure 4: (a) Transmission of waves through a multilayer realization consisting of silver-tio2 layers each of 10 nm thickness (total 16 layers) including loss and dispersion. The high-$k$ modes are seen to tunnel across the layers. (b) In the effective medium limit there are infinitely many high-$k$ waves in the type I and type II HMM. Notice there is no upper cut-off in the effective medium limit but in the multilayer realization the size of the unit cell imposes a strict upper limit to tunneling. ($k_0$ is the free space wavevector)

4. Applications

- Fluorescence engineering

The presence of high-k waves opens a new route into which quantum emitters can decay when placed on an HMM substrate. In the near field of any medium, there are in general three routes of decay corresponding to the types of electromagnetic modes supported by the structure. A quantum emitter or fluorophore can emit into propagating waves of vacuum or bound modes (such as waveguide modes or surface plasmon polaritons) and if the body is absorptive ( $\text{Im}(\varepsilon) \neq 0$ ), the third non-radiative route for relaxation is opened up. Fluorescence can be completely quenched due to near-field absorption.

The physical origin for quenching can be understood by considering the size of the fluorescent dye molecule (point dipole like) which is far below the size of the wavelength. This implies waves with large wavevectors are necessarily emitted by the emitter which do not normally carry energy to the far-field. However these waves can be completely absorbed in the near field by lossy structures. In stark contrast to quenching, the HMM couples to these waves with large wavevectors leading to radiative relaxation in the near field. The signature of this coupling is the enhanced decay rate of dye molecules placed on hyperbolic metamaterial substrates. The coupling to high-$k$ states opens the possibility of high contrast fluorescence imaging as well as sensitive phase measurements [20,21,44]. The decay rate in the near field at a distance 'd' from the HMM is dominated by high-$k$ waves and for type I HMM is given by

$$\Gamma_{high-k} \approx \frac{\mu_\perp^2}{8\hbar d^3} \frac{2\sqrt{|\varepsilon_{xx}|\varepsilon_{zz}}}{1+|\varepsilon_{xx}|\varepsilon_{zz}} \qquad (2)$$

where $\mu_\perp$ is the dipole moment of a perpendicularly oriented dipole. The decay rate into the high-*k* modes of the HMM can exceed the rate of emission into vacuum by factors of 10 even without any subwavelength confinement of the emitters.

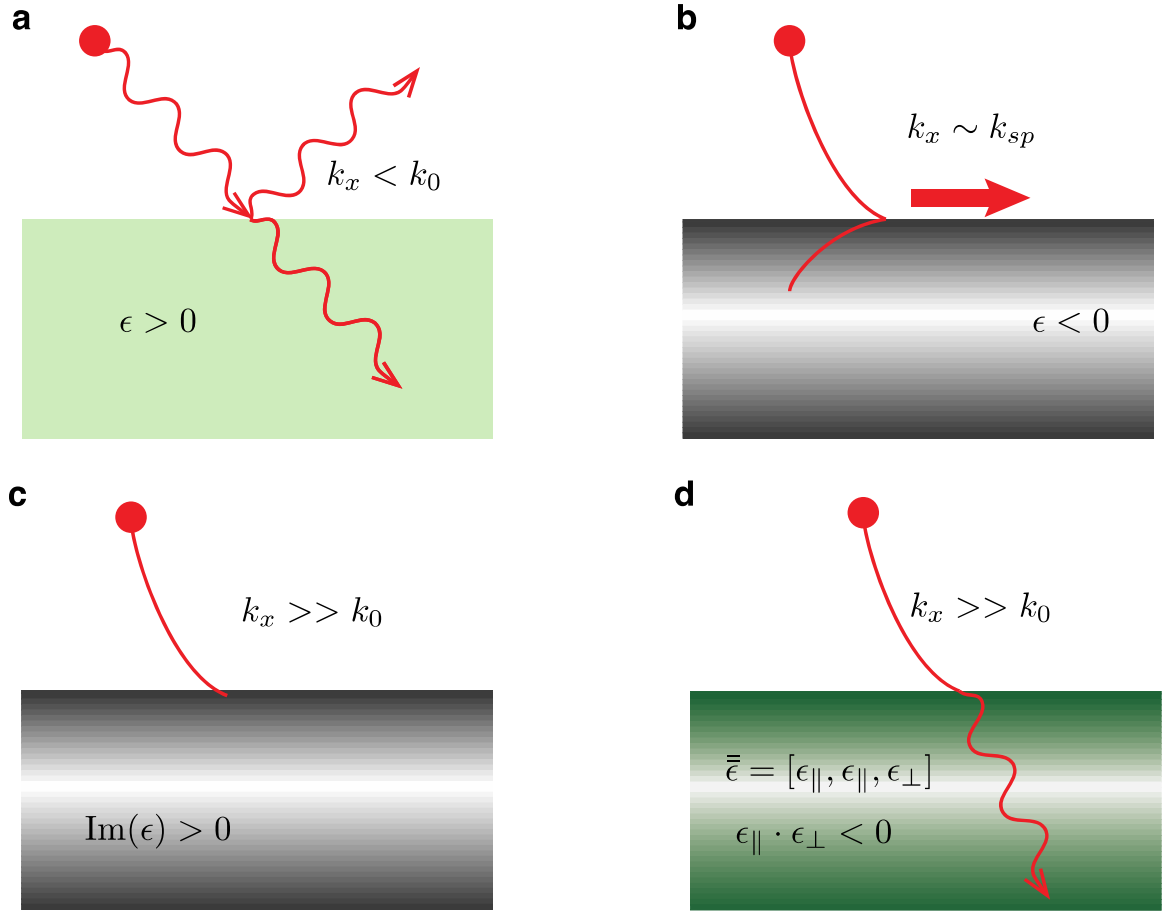

Figure 5. A subwavelength emitter such as quantum dot emits light with all spatial frequencies (i.e. wavevectors). (a) The light can couple to propagating waves in vacuum which are reflected and transmitted by a slab (eg. dielectric) placed in the vicinity of the emitter (b) If the slab is metallic, another route opens up which corresponds to the coupling of light to surface plasmons. The wavector of light that couples to the surface plasmon polariton is $k_{sp} > k_0$. (c) The light with large wavevectors emanating from the emitter cannot propagate in vacuum. If the slab is lossy (metal or dielectric) these high-*k* waves are simply absorbed. This is the phenomenon of quenching occuring in the near field of any lossy object. (d) A slab of hyperbolic metamaterial allows the propagation of these high-k waves. In the near field, the subwavelength emitter couples most efficiently to the HMM states as compared to plasmons or propagating waves in vacuum. This is due to the availability of a large number of these HMM states.

The validity of the effective medium theory and point dipole approximation for spontaneous emission has been studied in detail [19,35,36,45].

- Super-planckian thermal emission

The above mentioned high-$k$ states which lead to decrease in radiative spontaneous emission lifetime can also play a key role in thermal conductivity [32] and thermal emission [30]. This was initially pointed out in multiple references [19–21,31,32] as well as [46,47]. Due to the enhanced density of states, a hyperbolic metamaterial in equilibrium at temperature T emits super-planckian thermal radiation [30]. This can lead to near field thermal energy transfer beyond the black body limit. The main property that sets it apart from other resonant methods of super-planckian thermal emission is the broad bandwidth and also presence of topological transitions [30].

In the effective medium limit, the thermal energy density in the near field of the hyperbolic metamaterial is given by

$$u(z,\omega,T)^{z\ll\lambda} \approx \frac{U_{BB}(\omega,T)}{8}\left[\frac{2\sqrt{|\varepsilon_{xx}\varepsilon_{zz}|}}{(k_0 z)^3(1+|\varepsilon_{xx}\varepsilon_{zz}|)} - \varepsilon'' \frac{2(\varepsilon_{xx}+\varepsilon_{zz})}{(k_0 z)^3(1+|\varepsilon_{xx}\varepsilon_{zz}|)^2}\right] \quad (3)$$

where $u(z,\omega,T)$ is the energy density at a distance z and frequency $\omega$ while T denotes the temperature. $k_\rho = \sqrt{k_x^2+k_y^2}$, $k_0 = \omega/c$, $\varepsilon_\parallel = \varepsilon_{xx} + i\varepsilon''$, $\varepsilon_\perp = \varepsilon_{zz} + i\varepsilon''$ and $\varepsilon_{xy}\cdot\varepsilon_{zz} < 0$ and $U_{BB}(\omega,T)$ is the black body emission spectrum at temperature T. These results were recently presented [30], where the fluctuational electrodynamics of hyperbolic metamaterials was developed. This has many implications for thermal engineering using metamaterials and experiments are currently underway using phonon-polaritonic HMMs [43] to measure this unique effect using topological transitions [30,44].

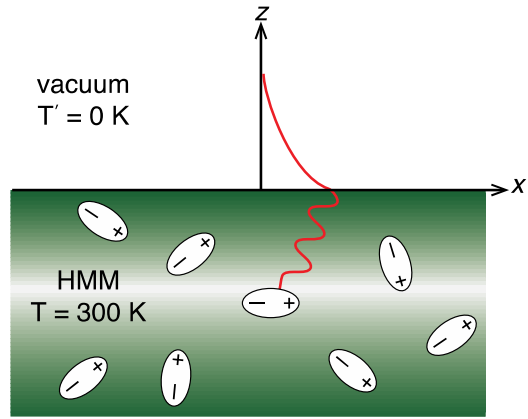

Figure 6: Recently, a fluctuational electrodynamics of hyperbolic metamaterials was developed and super-planckian thermal emission from these media was predicted [30]. In the near-field, a heated HMM can emit and thus transfer thermal radiation beyond the black body limit. This paves the way for near-field thermal engineering using metamaterials.

- Nano-imaging

The hyperlens is an imaging device made of hyperbolic metamaterials which can break the far-field diffraction limit [23,25]. The subwavelength resolution in the far-field arises from the cylindrical (or spherical) nature of the HMM. The device can be fabricated by the same multilayer design principles presented earlier. In the future, integrating the hyperlens with microfluidic channels can make a significant impact for real time bio-imaging.

The functioning of the device can be understood by noting that the diffraction limit arises since waves with large spatial frequencies which carry subwavelength information are evanescent in vacuum. This leads to a loss of information in the far field image. Near the HMM, these high-k waves are captured and turned into propagating waves and the information is carried within the HMM. Conservation of angular momentum ($m \sim k_\theta r$) ensures that at the outer surface of the hyperlens the tangential momentum is reduced and the waves can escape to the far-field. Another interpretation of far-field imaging is based on the realization that hyperbolic media only allow waves to propagate along subwavelength resonance cones in the radial direction [48,49]. Thus points on the inner surface are mapped one-to-one to points on the outer surface. Thus the spacing between diffraction-limited points on the inner radius can be much larger than the wavelength on the outer surface, if the outer radius is chosen to be big enough [49]. These well separated points can be viewed by a conventional microscope even if the original points cannot.

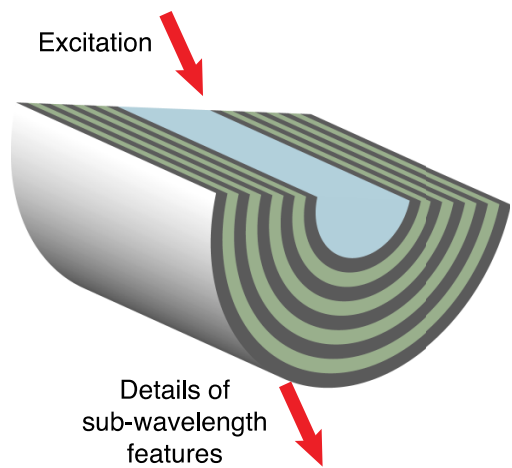

Figure 7: Integrating microfluidic channels with the hyperlens can lead to real-time super-resolution imaging for biological applications. The schematic shows the optofluidic channel and the cylindrical multilayer hyperlens.

- Sub-surface sensing

A doped semiconductor based HMM can be used for non-invasive sub-surface sensing [26] and sub-diffraction imaging in the IR. The semiconductor InGaAs can be doped to have its metallic properties tuned across the IR. Therefore, a multilayer structure consisting of InGaAs/AlInAs layers can be constructed to have a broadband ($\Delta\lambda \sim 2$ µm) Type 1 HMM behavior across the IR [16]. AlInAs acts as the dielectric. Potential applications and device geometries are described below

Sub-surface sensing and imaging are especially important for IR fingerprinting. Combining IR spectroscopy with sub-diffraction imaging may allow for chemical identification on a truly molecular scale. Another application is quality control in intergrated circuits (ICs) or micro-electro-mechanical devices (MEMs) since silicon is transparent to IR wavelengths. Therefore, cracks or defects in ICs or MEMs can be detected through the scattering and diffraction of IR light at these defects. The main problem with the conventional technique is the diffraction limit: cracks and defects smaller than the illuminating wavelength can not be imaged non-invasively. In terms of IR this corresponds to imperfections of the size ~2-3µm. By using a HMM, sub-diffraction resolution may be obtained so that defects with deep sub-wavelength dimensions may be observed non-invasively.

The applications described above are conceptualized in figure 8. Refractive index changes may be detected non-invasively through HMM slab. The emission power densities are overlayed to demonstrate the capability of sub-surface sensing. The power densities are calculated using the Green's Function approach for an effective medium HMM slab.

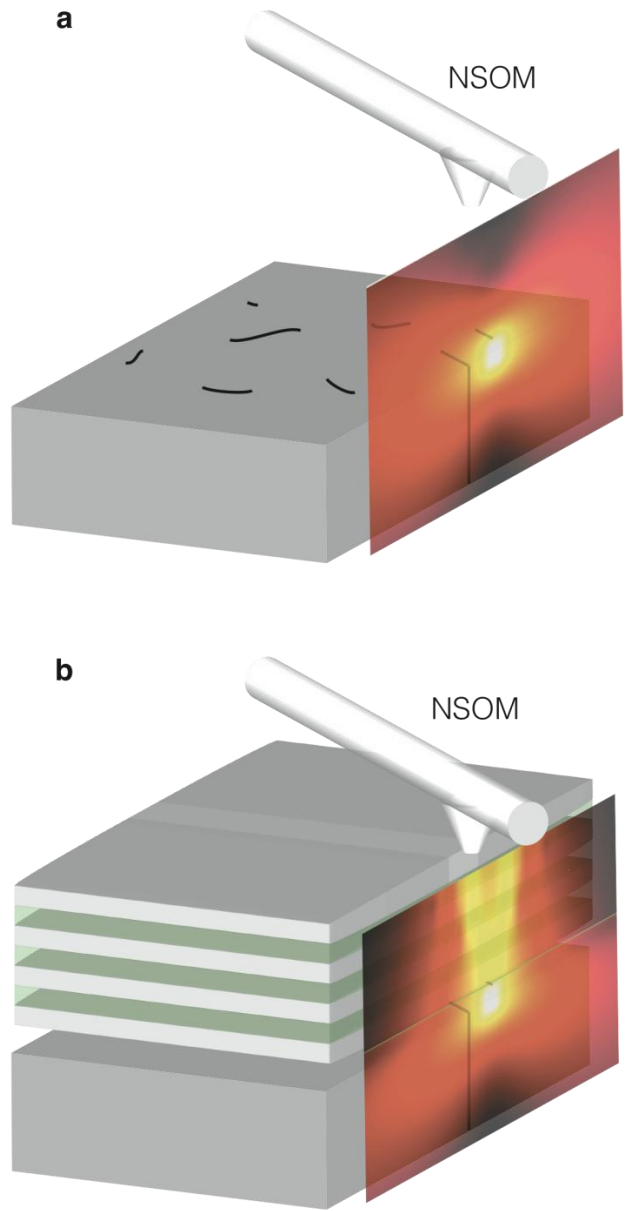

Figure 8 *Sub-Surface Sensing*: Illumination of a Si wafer from below by λ ~ 9.8 μm. The incident light scatters through cracks in the wafer. (a) A sub wavelength dimension crack in a Si wafer can be detected by a near field scanning optical microscope (NSOM) only if the detector is brought very close to the crack (<λ/6) (b) If a AlInAs/InGaAs HMM is placed above the wafer, then the NSOM can detect the crack up to 2-3 × λ from the wafer. The HMM allows high spatial frequencies to propagate and enhances their intensity in the image.

- Dyakonov plasmons

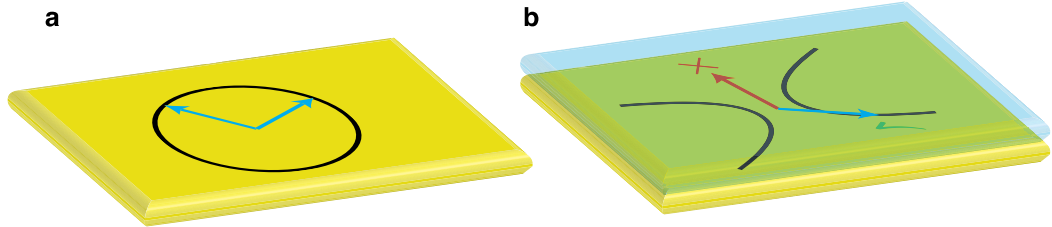

Figure 9. Dyakonov Plasmonics: Schematic showing (a) isofrequency curve of conventional plasmons on an isotropic gold film. The circular isofrequency curve corresponds to the fact that plasmons can propagate in all directions (blue arrows). (b) Isofrequency curve becomes hyperbolic when an anisotropic slab is placed on top of the gold film. This occurs in the range of frequenies when conventional plasmons which occur along the high symmetry directions are allowed in one direction (blue arrow) but are not allowed in the perpendicular direction (red arrow).

Hyperbolic isofrequency curves can also be obtained for two-dimensional surface states. This is schematically shown in figure 9. The isofrequency curve for in-plane surface plasmon polaritons on a gold substrate is circular. As opposed to this, when a uniaxial crystal is placed on top of the metal film with different indices along the principle axes there exists the possibility that plasmons are allowed to propagate in one direction but not in the perpendicular direction. This gives rise to in-plane hyperbolic isofrequency curves for the surface plasmon polaritons. They are a mixed state consisting of both TE and TM polarized light and are called Dyakonov plasmons [28].

The above example is that of interface states on a metallic substrate with a uniaxial dielectric on top. Hyperbolic metamaterials can support surface states with similar unconventional properties [28]. If the principle axis of the hyperbolic metamaterial (z-axis) lies in the interface plane with vacuum, then Dyakonov plasmon solutions with hyperbolic isofrequency curves are allowed. Dyakonov plasmons show all the properties of bulk hyperbolic states such as subdiffraction imaging and directional resonance cone beaming [27,28,50].

Note that the above mention states are fundamentally different from conventional surface plasmon polariton (SPP) solutions allowed in type II metamaterials when the interface plane permittivity is negative and the principle axis (z axis), normal to the plane, has positive dielectric permittivity. All modes of hyperbolic media (high-k, SPP and Dyakonov plasmons) are tunable since the optical constants governing them depend on the metallic fill fraction.

5. Future and Conclusion

The last decade has seen tremendous progress in the physics and nanofabrication of various classes of metamaterials. The next decade is set for metamaterial applications in different fields. For devices in the visible and near-IR wavelength ranges, hyperbolic metamaterials are expected to lead the way due to their varied properties and applicability. One major direction of application will be quantum nanophotonics [51,52]. We have shown here the potential of HMMs for nano-imaging, subsurface sensing, Dyakonov plasmonics, fluorescence engineering and thermal emission control. This paper should help experimentalists gather a unified view of the multiple applications of hyperbolic metamaterials for designing devices.

Acknowledgements

Z. Jacob wishes to acknowledge E. E. Narimanov and S. Pramanik for fruitful discussions. This work was partially supported by the National Science and Engineering Research Council of Canada, Canadian School of Energy and Environment, Alberta Nanobridge and Alberta Innovates.